\documentclass[authoryear,preprint,12pt,3p]{elsarticle}

\usepackage{amssymb}

\usepackage{lineno}

\newcommand*\patchAmsMathEnvironmentForLineno[1]{%
  \expandafter\let\csname old#1\expandafter\endcsname\csname #1\endcsname
  \expandafter\let\csname oldend#1\expandafter\endcsname\csname end#1\endcsname
  \renewenvironment{#1}%
     {\linenomath\csname old#1\endcsname}%
     {\csname oldend#1\endcsname\endlinenomath}}%
\newcommand*\patchBothAmsMathEnvironmentsForLineno[1]{%
  \patchAmsMathEnvironmentForLineno{#1}%
  \patchAmsMathEnvironmentForLineno{#1*}}%
\AtBeginDocument{%
\patchBothAmsMathEnvironmentsForLineno{equation}%
\patchBothAmsMathEnvironmentsForLineno{align}%
\patchBothAmsMathEnvironmentsForLineno{flalign}%
\patchBothAmsMathEnvironmentsForLineno{alignat}%
\patchBothAmsMathEnvironmentsForLineno{gather}%
\patchBothAmsMathEnvironmentsForLineno{multline}%
}

\usepackage{graphicx}
\usepackage{subfigure}
\usepackage{float}
\usepackage{amsmath}
\usepackage{amsfonts}
\usepackage{subfigure}
\usepackage{color}
\usepackage{multirow}
\usepackage{xcolor}

\definecolor{mycolor}{RGB}{0,0,0}  

\journal{Communications in Nonlinear Science and Numerical Simulation}

\begin{document}

\begin{frontmatter}



\title{An artificial neural network framework for reduced order modeling of transient flows}



\author{Omer San}\corref{cor1}
\cortext[cor1]{Corresponding author.}
\ead{osan@okstate.edu}

\author{Romit Maulik}
\author{Mansoor Ahmed}

\address{School of Mechanical and Aerospace Engineering, Oklahoma State University, Stillwater, Oklahoma 74078, USA}

\begin{abstract}
This paper proposes a supervised machine learning framework for the non-intrusive model order reduction of unsteady fluid flows to provide accurate predictions of non-stationary state variables when the control parameter values vary. Our approach utilizes a training process from full-order scale direct numerical simulation data projected on proper orthogonal decomposition (POD) modes to achieve an artificial neural network (ANN) model with reduced memory requirements. This data-driven ANN framework allows for a nonlinear time evolution of the modal coefficients without performing a Galerkin projection. Our POD-ANN framework can thus be considered an equation-free approach for latent space dynamics evolution of nonlinear transient systems and can be applied to a wide range of physical and engineering applications. Within this framework we introduce two architectures, namely sequential network (SN) and residual network (RN), to train the trajectory of modal coefficients. We perform a systematic analysis of the performance of the proposed reduced order modeling approaches on prediction of a nonlinear wave-propagation problem governed by the viscous Burgers equation, a simplified prototype setting for transient flows. We find that the POD-ANN-RN yields stable and accurate results for test problems assessed both within inside and outside of the database range and performs significantly better than the standard intrusive Galerkin projection model. Our results show that the proposed framework provides a non-intrusive alternative to the evolution of transient physics in a POD basis spanned space, and can be used as a robust predictive model order reduction tool for nonlinear dynamical systems.
\end{abstract}


\end{frontmatter}


\section{Introduction}
\label{sec:intro}

The full resolution of the dynamics of engineering flows requires a computational expense that is as yet infeasible for any practical purpose. Due to these overwhelming computational demands, model order reduction type approaches are gaining popularity. They are extensively used for many optimization and control applications \citep{akhtar2015using,buffoni2006low,dyke1996modeling,fang2009pod,fortuna2012model,freund1999reduced,noack2011reduced,lucia2004reduced,roychowdhury1999reduced,silveira1996efficient}. Among the multitude of reduced order modeling approaches, \textit{proper orthogonal decomposition} (POD) has emerged as a popular technique for the study of dynamical systems \citep{aubry1988dynamics,holmes1998turbulence,ly2001modeling,kerschen2005method,cizmas2008acceleration,amsallem2012stabilization,taira2017modal}. To clarify, POD is also documented under terminology such as the Karhunen-Lo\`{e}ve expansion \citep{loeve1955probability}, principal component analysis \citep{hotelling1933analysis} or empirical orthogonal function \citep{lorenz1956empirical}. POD extracts the most energetic modes from a collection of high fidelity numerical simulations of the governing equations of the dynamic system being studied. These bases are then used to reduce the degrees of freedom of the governing equation to scales that are computationally tractable. In particular, the fluid mechanics community has traditionally used POD as a method of extraction of the large scale coherent structures (represented by these modes) for the purpose of statistical pattern recognition \citep{lumley1967structures}. This is because the global POD modes optimally span our physical space through a considerably truncated number of bases to resolve attractors and transients well.   

\textcolor{mycolor}{The evolution equations for the lower order system are then obtained using the Galerkin projection (GP) method.} The POD-GP method has been extensively used to provide fast and accurate simulations of large nonlinear systems  \citep{berkooz1993proper,kunisch2002galerkin,lucia2004reduced,borggaard2007interval,weller2010numerical,benner2015survey,brunton2015closed,taira2017modal,rowley2017model}. 
The Galerkin projection is devised so as to generate a reduced model where the truncated linear coefficients of the POD bases are evolved through time as a set of ordinary differential equations (ODEs). This process is denoted a \textit{projection} since it is obtained by an orthogonal projection of the governing equations onto the truncated POD modes. The result is a considerably truncated system which continues to represent the most significant characteristics of the governing equations. The small memory requirement and low computational expense of this framework make it particularly suitable for post-processing analysis, optimization and control type problems where repeated model evaluations are required over a large range of parameters. This approach is significantly effective for quasi-stationary, time-periodic and decaying problems but might be challenging for highly non-stationary, convective and nonlinear problems. 

The drawbacks are generally due to the lack of adaptation of the POD modes as well as the loss of information due to the truncation being limited to the most energetic of modes \citep{cordier2013identification,el2016new}. Amongst the various efforts used to mitigate these drawbacks include modeling the effect of truncated modes on the retained ones, an approach known as closure modeling \citep{wang2011two,borggaard2011artificial,wang2012proper,osth2014need,san2015stabilized,wells2017evolve,xie2017approximate,xie2018numerical,xie2018data}, and developing strategies to construct a more representative basis \citep{iollo2000stability,buffoni2006low,bui2007goal,kunisch2010optimal,carlberg2011low,san2015principal,ahmed2018stabilized}. On the other hand, the energy balance and mass conservation properties of projection based reduced order models have also been studied for fluid flows \citep{mohebujjaman2017energy}. \textcolor{mycolor}{However, the projection-based model reduction approaches have limitations especially for complex systems such as general circulation models, since there is a lack of access to the full-order model (ROM) operators or the complexity of the forward simulation codes that renders the need for obtaining the full-order operators \citep{peherstorfer2016data}. Therefore, there is a recent interest in generating fully non-intrusive approaches without the need for access to full-order model operators to establish surrogate models \citep{audouze2013nonintrusive,mignolet2013review,xiao2015non,xiao2015non,hesthaven2018non,hampton2018practical,chen2018greedy,xiao2019domain,wang2019non}.
}

Alternatively, machine learning offers promise to generate accurate parametric reduced order models. \textcolor{mycolor}{There are a broad range of opportunities for the development of new deep learning architectures for reduced order models and numerical analysis of these machine learning approaches in applications of challenging flows.}  Recently, a supervised machine learning approach has been used to develop feasible regions in the parameter space where the admissible target accuracy is achieved with a predefined reduced order basis \citep{moosavi2015efficient}. Neural networks based closure schemes have also been developed by the authors to compute a stabilization term (i.e., eddy viscosity) for Galerkin projection based reduced order models \citep{san2018extreme,san2018neural,san2018machine}. 
\textcolor{mycolor}{A deep residual recurrent neural network has been introduced as an efficient model reduction technique for nonlinear dynamical systems \citep{kani2017dr}. In a context similar to this recurrent neural network architecture, we formulate our problem in low-dimensional embedded structures that capture the major portion of the energy of the system in order to take advantage of lower-dimensional coherent structures that are responsible for the bulk mass, momentum, and energy transfer instead of performing a very large-scale simulation that considers all structures. A recent investigation of a spectral POD (SPOD) spanned model order reduction framework for fluid flows has been performed by \cite{lui2019construction} using the deep feedforward neural networks. A long short-term memory (LSTM) architecture based model reduction approach has also been recently introduced to complement an imperfect reduced order model with data-streams utilizing a regularized hybrid framework \citep{wan2018data}.}

The present study aims to bypass the Galerkin projection through the use of a single-layer artificial neural network (ANN) by introducing two different architectures, namely a sequential network (SN) and a residual network (RN). ANNs can be classified under the broader category of machine learning methods (i.e., systems that learn from data) and are a mathematical representation of the biological neural networks found in the human central nervous system. They have been used for a wide variety of applications such as function approximation, classification, data processing and dynamic systems control \citep{widrow1994neural}. Briefly, an ANN can be used to setup a nonlinear relationship between a desired set of inputs and targets provided a large amount of benchmark data for the underlying relationship is available. This fitting to available data (also known as \textit{training}) ideally generates a set of linear combinations of parameters and transfer functions which replicate the mean behavior of the underlying phenomena. This allows for the representation of subtle relationships which cannot be expressed explicitly in a functional form. The interested reader is directed to the excellent introductory text by \cite{demuth2014neural} on the development of a variety of ANN architectures and their underlying principles. We stress here that generating a robust non-intrusive reduced order modeling (ROM) approach is highly desirable for its ability to preclude model-form uncertainty in our forward simulations for reduced computational degree of freedom and is thus a highly active area of research \citep{xiao2015non1,xiao2015non2,bistrian2015improved,peherstorfer2016data,lin2017non,kramer2017sparse,bistrian2017randomized}. In fact, several recent studies have investigated the suitability of ANN variants for fully non-intrusive dynamics capture in latent space \citep{mohan2018deep,wan2018data,wang2018model}. 

With regard to the ROM methodology, we implement our framework in a context similar to the works of \cite{narayanan1999low} and \cite{khibnik2000analysis} where the POD-GP approach was replaced by an ANN-based evolution of the temporal dynamics of the ROM for the purpose of active separation control in a planar diffuser. Their dataset for ANN training was obtained both from direct numerical simulations of the Navier-Stokes equations as well as experiments. In \cite{sahan1997artificial}, an ANN was used to devise a feedback control circuit for a transitional flow in a grooved channel. The dataset for ANN training was obtained through a POD-GP simulation itself. Both works demonstrated the unique advantages of the ANN approach, for example, the reduction of a system of governing PDEs to a no-equation dynamic system through data-driven learning. We note that ANNs also find great utilization in the field of feedback flow control where they are used to generate a direct mapping of flow measurements to actuator control systems \citep{gillies1995low,gillies1998low,gillies2001multiple,faller1997unsteady,hocevar2004experimental,efe2004modeling,efe2005control,lee1997application}. General machine learning based methods (of which ANN is a subset) are also growing in popularity for fluid flow based applications and represent a computationally viable alternative to the full Navier-Stokes equations \citep{bright2013compressive,gautier2015closed,muller1999application}. An excellent source of information about some viable machine learning classes can be found in \cite{ling2015evaluation}. In turbulence modeling, ANN approaches have also been used in combination with reduced degree-of-freedom implementations of the Navier-Stokes framework, for instance in the quantification of errors in functional closure models \citep{singh2016using} and for data-driven closure modeling \citep{milano2002neural,tracey2013application,ling2016machine,ling2016reynolds,maulik2017neural,gamahara2017searching}. 

This paper develops a framework for generating an ANN in order to create a very fast integrator for certain types of partial differential equation (PDE) systems (e.g., one-dimensional Burgers equation in our demonstrations). The ANN is created to represent a map for the trajectory of the POD basis coefficients evolving from time $n$ to time $n+1$. The projection of snapshot data is used to \emph{train} the ANN, i.e., make it choose the weights to best fit the data in the POD spanned reduced order space. Therefore, this model reduction approach is denoted as PON-ANN in our study. \textcolor{mycolor}{Using only snapshots of the state variables, our data-driven approach can thus be considered truly non-intrusive, since any prior information about the underlying governing equations is not required for generating the reduced order model.} This idea can be considered as a competitor to the POD-GP, and the advantages are that this is an equation-free method, and also it can be more accurate since it can be trained with more data. Although many new model order reduction techniques perform better than standard POD-GP model, as highlighted above, we focus here on comparing the proposed approach with the POD-GP model only. Future comparative studies should be carried out to explore the feasibility of such non-intrusive models. Before proceeding with our discussion, we describe the major questions we wish to answer in this work:
\begin{itemize}
  \item Can the POD-ANN non-intrusive ROM approach be used for transient dynamical systems such as nonlinear Burgers equation?
  \item How does an ANN-based evolution of the modal coefficients compare to the POD-GP approach?
  \item Can a trained POD-ANN (using a particular dataset) be used to interpolate or extrapolate to flow problems with slight differences in temporal evolution and physics (i.e., for different control parameter values such as Reynolds number)?
  \item \textcolor{mycolor}{Which POD-ANN architecture is more stable and robust in forecasting beyond the training data range?} 
\end{itemize}
To answer these questions, we investigate a test case given by the nonlinear viscous Burgers equation to compare both the POD-ANN and POD-GP approaches.

\section{Mathematical modeling}

\begin{figure}[!ht]
\centering
\includegraphics[width=0.9\textwidth]{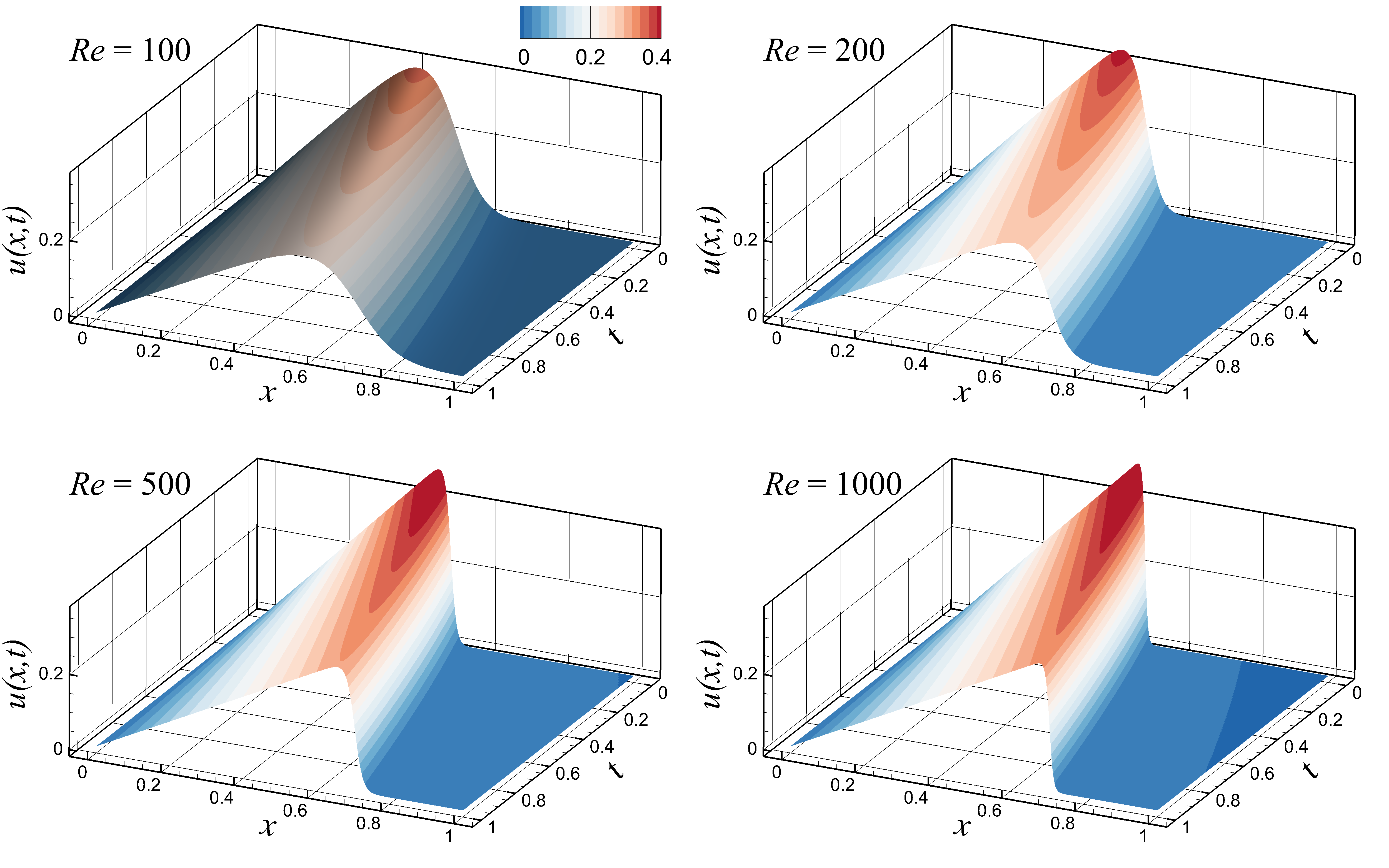}
\caption{Space-time solution of the Burgers equation for various Reynolds numbers within our database. Note that both the POD analysis and ANN training consider a database constructed by using 10 equally spaced values of Reynolds number between $Re=100$ and $Re=1000$, i.e., $Re = [100, 200, ..., 1000]$.}
\label{Fig:burger1}
\end{figure}

The viscous Burgers equation is used in our study as a test case to explain the proposed model reduction framework. In its original form, the Burgers equation can be written as
\begin{equation}
\label{e:1a}
    \frac{\partial u}{\partial t} + u \frac{\partial u}{\partial x} = \frac{1}{Re} \frac{\partial^2 u}{\partial x^2} \, , \quad x \in [0,1], \quad 
    t \in [0,1]\, ,
\end{equation}
where $Re$ is a non-dimensional Reynolds number. We must mention here that this equation is generally considered a framework for preliminary evaluation of numerical methods for analysis of fluid flow applications as it possesses the hallmarks of general nonlinear multidimensional advection-diffusion problems. We consider a convective system in our model order reduction assessments since it characterizes localized flow structures such as shock waves. The above PDE can be solved exactly to obtain an analytical formulation for time evolution of the the field variable $u(x,t)$ given by \citep{maleewong2011line}
\begin{align}
\label{eq:exact}
u(x,t) = \frac{\frac{x}{t+1}}{1+\sqrt{\frac{t+1}{t_0}}\exp(Re \frac{x^2}{4t+4})},
\end{align}
where $t_0 = \exp(\frac{Re}{8})$. This exact expression is used to generate snapshot data for our forthcoming model order reduction analysis. The database is constituted with Eq.~(\ref{eq:exact}) using $N_x = 1024$ spatial collocation points at each snapshot. Our database consists of snapshots obtained  by using 10 equally spaced values of Reynolds number, i.e., $Re = [100, 200, ..., 1000]$.  Figure~\ref{Fig:burger1} shows the space-time behavior of the four representative solutions at different Reynolds number.  

\section{Proper orthogonal decomposition}
\label{sec:ROM}

In this section we explore the POD approach for reduced order modeling to capture unsteady, convective and non-periodic dynamics of the underlying governing partial differential equations. We elaborate the procedure for calculating the orthonormal bases and corresponding coefficients for the Burgers problem and note that an extension to higher dimensions is straightforward. For a detailed discussion on this technique we refer the reader to \cite{san2013proper}.

\begin{figure}[!ht]
\centering
\mbox{
\includegraphics[width=0.9\textwidth]{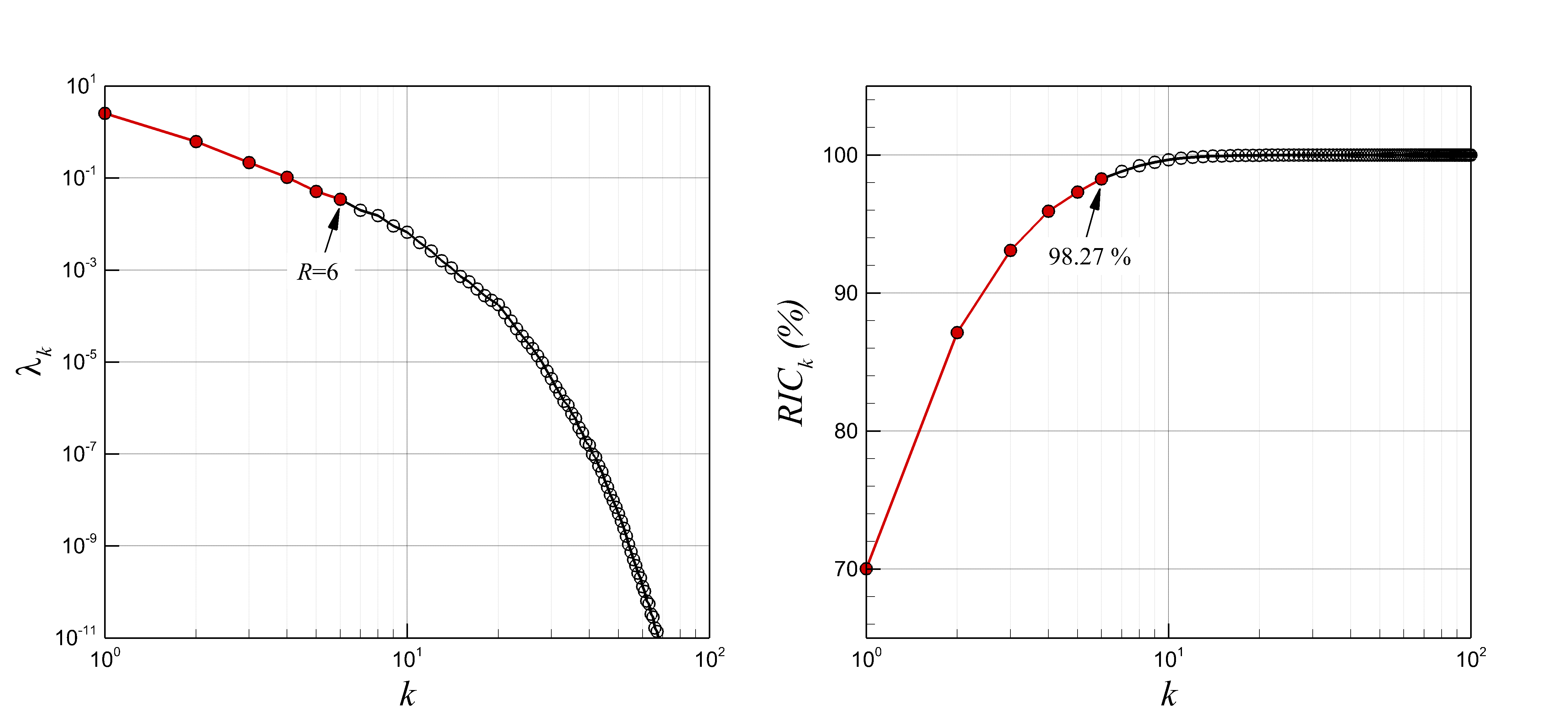}
}
\caption{Eigenvalues of the snapshot correlation matrix (left) and the corresponding relative information content (RIC) index (right) for the space-time solution of the viscous Burgers equation. Note that we retain only the most energetic 6 POD modes in our model order reduction analysis.}
\label{Fig:burger2}
\end{figure}

\begin{figure}[!ht]
\centering
\mbox{
\includegraphics[width=0.9\textwidth]{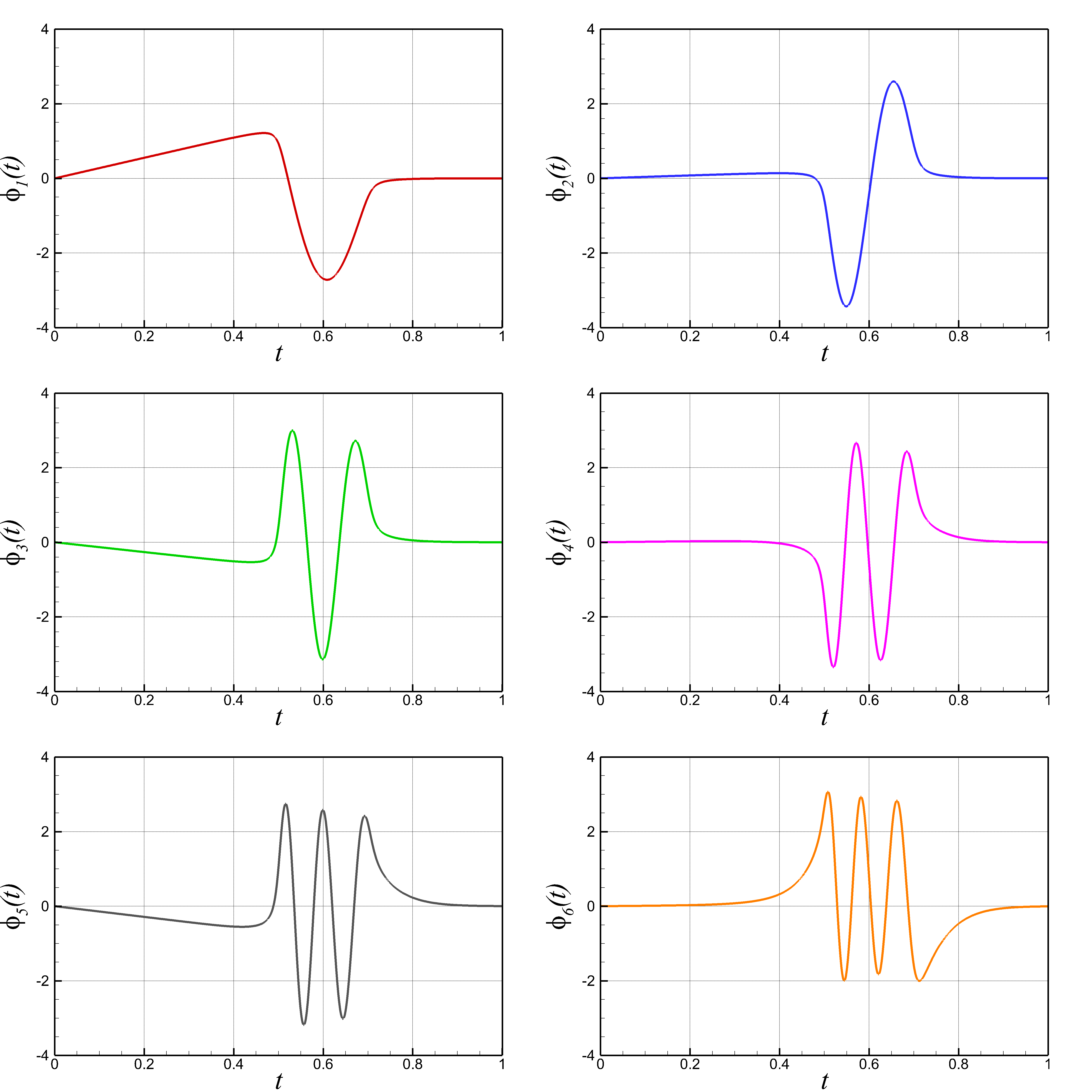}
}
\caption{Illustration of the most energetic POD basis functions generated using a total of 1010 data snapshots from $Re = 100$ and $Re = 1000$.}
\label{Fig:burger3}
\end{figure}

A representative POD basis can be constructed from any arbitrary scalar field variable $f$ at different times (also known as snapshots). These snapshots are generally obtained by solving the governing equations we are attempting to model using a regular full-order model (FOM) approach. The numerical methods used to obtain this field data are explained in Section \ref{sec:num}. In the following, we will utilize the index $n$ to indicate a particular snapshot in time. For the POD approach we utilize a total of $N$ snapshots for the field variable, i.e., $f(\textbf{x},t^{(n)})$ for $n=1,2,...,N$. The flow field data thus obtained is first decomposed into the mean and fluctuating part as follows
\begin{align}\label{POD_1}
  f(\textbf{x},t) = \bar{f}(\textbf{x}) + f'(\textbf{x},t),\ \ \ \ \bar{f}(\textbf{x}) = \frac{1}{N} \sum_{n=1}^{N} f(\textbf{x},t^{(n)}),
\end{align}
where $\bar{f}$ implies a temporal averaging of a particular point value in the field and $f'$ contains the fluctuating quantity. We must remark here that $\bar{f}$ is a function of space alone whereas $f'$ is a function of both space and time. A correlation matrix may be constructed using the fluctuating components of the snapshots to give
\begin{align}\label{POD_2}
  C_{ij} = \int_{\Omega} f'(\textbf{x},t^{(i)}) f'(\textbf{x},t^{(j)}) d\textbf{x},
\end{align}
where $\Omega$ is the entire spatial domain and $i$ and $j$ refer to the $i^{th}$ and $j^{th}$ snapshots. The time correlation data matrix $C$ is a non-negative symmetric square matrix of size $N \times N$. If we define the inner product of any two fields $f_1$ and $f_2$ as
\begin{align}\label{POD_3}
  (f_1,f_2) = \int_{\Omega} f_1(\textbf{x}) f_2(\textbf{x}) d\textbf{x}
\end{align}
we may express the correlation matrix as $C_{ij} = \left(f'(\textbf{x},t^{(i)}),f'(\textbf{x},t^{(j)})\right)$. In this study, we use the well-known Simpson's 1/3 integration rule for a numerical computation of the inner products. The optimal POD basis functions are obtained by performing an eigendecomposition of the $C$ matrix. This has been shown in detail in the POD literature (see, e.g., \cite{sirovich1987turbulence,holmes1998turbulence,ravindran2000reduced}). The eigenvalue problem can be written in the following form:
\begin{align}
\label{POD_4}
    CW=W\Lambda \, ,
\end{align}
where $\Lambda=\mbox{diag}[\lambda_1, \lambda_2, ..., \lambda_N ]$ is a diagonal matrix containing the eigenvalues of this decomposition and
$W$ =[$\boldsymbol w^{1}$, $\boldsymbol w^{2}$, ..., $\boldsymbol w^{N}$].
The eigenvalues are stored in descending order, $\lambda_1 \geq \lambda_2 \geq ... \geq \lambda_N$. Then the orthogonal POD basis functions can be written as
\begin{align}
\label{POD_5}
    \phi_1(\textbf{x}) = \sum_{n=1}^{N}w^{1}_{n}f'(\textbf{x},t^{(n)}), \quad \phi_2(\textbf{x}) = \sum_{n=1}^{N}w^{2}_{n}f'(\textbf{x},t^{(n)}), \quad ..., \quad \phi_N(\textbf{x}) = \sum_{n=1}^{N}w^{N}_{n}f'(\textbf{x},t^{(n)}) \, ,
\end{align}
where $w^{k}_{n}$ is the $n$th component of the $k$th eigenvector $\boldsymbol w^{k}$. Therefore, we emphasize that the POD modes are ranked according to the magnitude of their eigenvalue. The eigenvectors must also be normalized in order to satisfy the condition of orthonormality between bases:
\begin{align}
\label{POD_6}
    \left( \phi_i,\phi_j \right)=\left\{
                      \begin{array}{ll}
                        1, & i=j \\
                        0, & i\neq j \, .
                      \end{array}
                    \right.
\end{align}
It can be shown that, for Eq.~(\ref{POD_6}) to be true, the eigenvector $\boldsymbol w^{k}$ must satisfy the following equation:
\begin{align}
\label{POD_7}
\sum_{n=1}^{N}w^{k}_{n}w^{k}_{n}=\frac{1}{\lambda_k},
\end{align}
where $\lambda_k$ is the $k$th eigenvalue associated to the eigenvector $\boldsymbol w^{k}$. In practice, most of the subroutines for solving the eigensystem given in Eq.~(\ref{POD_4}) return the eigenvector matrix $W$ having all the eigenvectors normalized to unity. In that case, the orthogonal POD bases are given by
\begin{align}
\label{POD_8}
    \phi_k(\textbf{x}) = \frac{1}{\sqrt{\lambda_k}}\sum_{n=1}^{N}w^{k}_{n}f'(\textbf{x},t^{(n)})
\end{align}
where $\phi_k(\textbf{x})$ is the $k$th POD basis function. We remark here that the POD should be carried out independently for each field variable for multidimensional governing equations. We collect 101 snapshots between time $t=0$ and $t=1$ at each Reynolds number between $Re=100$ and $Re=1000$ using 10 different realizations (i.e., $Re = [100, 200, ..., 1000]$). Figure~\ref{Fig:burger2} shows the eigenvalues of the time correlation matrix $C$ and the corresponding relative information content (RIC) index, which is defined as \citep{gunzburger2012flow,tallet2016optimal}
\begin{align}
	RIC_{k} = \frac{\sum_{n = 1}^{k}\lambda_n}{\sum_{n=1}^{N}\lambda_n} \times 100,
\end{align}
where the number of snapshots is set to $N = 1010$ in our study. As shown in Table~\ref{tab:1}, the RIC index measures the percentage of captured energy for different number of POD modes. It is evident that a small number of modes are enough to capture the spatial and temporal dynamics of the system well. Therefore, we present our ROM analyses using $R=6$ largest (most energetic) POD modes capturing approximately 98.27 \% of the energy. Figure~\ref{Fig:burger3} illustrates the associated POD basis functions utilized in this study.

\begin{table}[!t]
\centering
\caption{The relative information content (RIC) index that measures the percentage of captured energy for different number of POD modes.}
\label{tab:1}
\smallskip
\begin{tabular}{p{1.5in}p{1in}}
\hline\noalign{\smallskip}
$R$ &   $RIC (\%)$ \\ \noalign{\smallskip} \hline\noalign{\smallskip}
  1   & 70.0273  \\
  2   & 87.1354  \\
  3   & 93.0921  \\
  4   & 95.9237  \\
  5   & 97.3327  \\
  6   & 98.2723  \\
  10   & 99.6722  \\
  20   & 99.9897  \\
\noalign{\smallskip}\hline
\end{tabular}
\end{table}

\section{Galerkin projection methodology}

The POD basis functions account for the essential dynamics of the underlying governing equations. After these data-driven empirical POD basis functions are obtained, a set of nonlinear ODEs can be derived by using a Galerkin projection. To build a projection based ROM, we truncate the system by considering the first $R$ largest POD basis functions with $R\ll N$. These POD modes correspond to the $R$ largest eigenvalues, $\lambda_1$, $\lambda_2$, ..., $\lambda_R$. Using these first $R$ largest POD basis functions the field variables are then approximated.


Figure~\ref{Fig:burger2} shows a collection of the modal values (i.e., the eigenvalues) of the POD-ROM analysis. The main motivation behind the construction of a POD-GP is evident when it is observed that the 6 most energetic modes retain 98.27\% of the total energy of the system. We note that these bases are all orthogonal to each other. For our test case, i.e., $u(x,t)=\bar{u}+u'(x,t)$, the Galerkin projection can be carried out in the following manner
\begin{align}
\label{e:14}
    u'(x,t)=\sum_{k=1}^{R}a_k(t)\phi_{k}(x) \, ,
\end{align}
where $a_k$ are the time dependent coefficients, and $\phi_{k}$ are the space dependent modes.
To derive the POD-ROM, we first rewrite the Burgers equation (i.e., Eq.~(\ref{e:1a})) in the following form
\begin{align}
\label{e:15}
    \frac{\partial u}{\partial t} = L[u] + N[u;u] \, ,
\end{align}
where $L[f]=\frac{1}{Re}\frac{\partial^2 f}{\partial x^2}$ is the linear operator and $N[f;g]= - f \frac{\partial g}{\partial x}$ is the nonlinear operator. We emphasize that the procedure can be easily extended to the general nonlinear PDEs (e.g., Boussinesq or Navier-Stokes equations). By applying this projection to our nonlinear system (i.e., multiplying Eq.~(\ref{e:15}) with the basis functions and integrating over the domain), we obtain the Galerkin POD-ROM, denoted POD-GP:
\begin{align}
\label{e:16}
    \left(\frac{\partial u}{\partial t},\phi_k\right) = (L[u],\phi_k) + (N[u;u],\phi_k), \quad \mbox{for} \quad k=1,2, ..., R \, .
\end{align}
Substituting Eq.~(\ref{e:14}) into Eq.~(\ref{e:16}), and simplifying the resulting equation by using the condition of orthonormality given in Eq.~(\ref{POD_6}), the POD-GP implementation can be written as follows:
\begin{align}
\label{e:17}
    \frac{d a_{k}}{dt} = B_{k} + \sum_{i=1}^{R} \mathfrak{L}_{ik} a_i + \sum_{i=1}^{R}\sum_{j=1}^{R} \mathcal{N}_{ijk}a_{i}a_{j}, \quad \mbox{for} \quad k=1,2, ..., R \, ,
\end{align}
where
\begin{eqnarray}
B_{k} &=&( L[\bar{u}] + N[\bar{u};\bar{u}],\phi_{k}), \label{e:18} \\
\mathfrak{L}_{ik} &=&(L[\phi_{i}] + N[\bar{u};\phi_{i}] + N[\phi_{i} ; \bar{u}],\phi_{k}), \label{e:20}  \\
\mathcal{N}_{ijk}    &=&( N[\phi_{i};\phi_{j}],\phi_{k}) \label{e:22} .
\end{eqnarray}

The POD-GP given by Eq.~(\ref{e:17}) consists of $R$ coupled ODEs and can be solved by a standard numerical method (such as the third-order Runge-Kutta scheme that was used in this study). The number of degrees of freedom of the system is now significantly lower. The vectors, matrices and tensors in Eqs.~(\ref{e:18})-(\ref{e:22}) are also precomputed quantities, which results in a dynamical system that can be solved very efficiently. To complete the dynamical system given by Eq.~(\ref{e:17}), the initial condition is given by using the following projection:
\begin{equation}
\label{e:23}
a_{k}(t=0)= \left( u(x,t=0)-\bar{u}(x),\phi_{k} \right) \, ,
\end{equation}
where $u(x,t=0)$ is the physical initial condition of the problem given in Eq.~(\ref{eq:exact}).

\section{Artificial neural networks}
\label{sec:ANN}

A simple feed-forward artificial neural network consists of $L$ layers with each layer possessing a predefined number of unit cells called neurons and is utilized for establishing nonlinear maps between two spaces of potentially different dimensionality through a supervised learning. Each of these layers may be considered an intermediate step in successive transformations between the input and output space and has an associated transfer function and each unit cell has an associated bias. Therefore, in our choice of network, any input to the neuron has a bias added to it followed by an activation through the transfer function. To describe this process using equations, we have \citep{demuth2014neural}
\begin{align}\label{ANN_1}
  s_i^l = \sum_j W_{ij}^l X_{j}^{l-1},
\end{align}
as the state of the signal in the $l^{th}$ layer receiving a set of inputs from the $(l-1)^{th}$ layer, where $W_{ij}^l$ stands for a matrix of weights linking the $l-1$ and $l$ layers with $X_j^{l-1}$ being the output of the $(l-1)^{th}$ layer. The output signal of the $l^{th}$ layer is now given by
\begin{align}\label{ANN_2}
  X_i^l = F(s_i^l + b_i^l),
\end{align}
where $b_i^l$ is the biasing parameter for the $i^{th}$ neuron in the $l^{th}$ layer. Every node (or unit cell) has an associated transfer function which acts on its input and bias to produce an output which is `fed forward' in the network. The nodes which take the raw inputs of our training data set (i.e., the nodes of the first layer in the network) perform no computation (i.e., they do not have any biasing or activation through a transfer function). The next layers are a series of unit cells which have an associated bias and activation function which perform computation on their inputs. These are called the hidden layers due to the indeterminate nature of their mathematical operations. The final layer in the network is that of the outputs. The output layer generally has a linear activation function with a bias which implies a simple summation of inputs incident to a unit cell with its associated bias. A mathematical description of this mapping operation can be expressed as
\begin{align}
	\mathbb{M} : \{p_1,p_2, \hdots, p_P\} \in \mathbb{R}^P \rightarrow \{q_1, q_2, \hdots, q_Q\} \in \mathbb{R}^Q.
\end{align}
where $P$ and $Q$ are the dimensions of the input and output spaces. In this investigation, we have used one hidden layer of neurons between the set of inputs and targets with a tan-sigmoid activation function. The tan-sigmoid function can be expressed as
\begin{align}\label{ANN_3}
  F(a) = \frac{2}{1+\exp(-2a)}-1.
\end{align}
The transfer function $F$ calculates the neuron's output given its net input. In theory, any differentiable function can qualify as an activation function \citep{zhang1998forecasting}, however, only a small number of functions which are bounded, monotonically increasing and differentiable are used for this purpose. The choice of using ANN is motivated by its excellent performance as a forecasting tool \citep{dawson1998artificial,kim2003nonlinear} and its general suitability in the machine learning and function estimation domain (e.g., see \cite{haykin2009neural} and references therein).

\subsection{POD transforms}
\textcolor{mycolor}{The key idea in our learning problem is the model reduction. Instead of posing our learning problem in a full-order space, we first apply projection to the full-order data and generate coefficients on a reduced subspace, and then construct the learning on this reduced-order space. Here we briefly describe the main procedure to generate such nun-intrusive reduced order models. 
Lets consider we have snaphots either coming from the full-order numerical simulations or experiments. Section 3 illustrates how to obtain representative POD modes. At any snapshot, $u^{(n)}$, our encoder operator can be written using the definition of inner product as 
\begin{align}
 a_{k}^{(n)} = \left(u^{(n)}-\bar{u},\phi_{k}\right), \quad \mbox{for} \quad k=1,2, ..., R,
\end{align}
where basis functions $\phi_{k}$ and the mean field $\bar{u}$ are already available. This can be interpreted as a forward POD transform from the FOM space to the ROM space. We then perform our supervised learning for $a_{k}^{(n)}$ to train a map for computing the modal coefficient trajectories. The details of the neural network architectures can be found in more detail in the following section. Finally, the decoder operator is defined as
\begin{align}
 u^{(n)} = \bar{u} + \sum_{k=1}^{R}a_{k}^{(n)}\phi_{k},
\end{align}
which can be referred to as an inverse POD transform. Indeed this expression can be used to generate full-order space data from the modal coefficients at any time.} 

\subsection{ANN training}
The training of a desired ANN is carried out by minimizing the error between the target and the inputs to determine a set of best fit parameters. This is also known as supervised learning where labeled data is utilized for gradient based optimization. Our best fit parameters obtained from optimization are biases and linear weights that have captured the underlying relationship between the targets and inputs and may now be used to predict target data for inputs a-posteriori. The main advantage of using the ANN approach over traditional statistical regression models is that comparatively smaller data sets for training are suitable. We shall describe the individual features of the ANN architecture relevant to our test case in the sections that follow.

The training of an ANN architecture may be undertaken by a variety of optimization algorithms. \textcolor{mycolor}{For our purpose, we have employed the Bayesian regularization minimization algorithm which is one of the popular approaches to ANN training within the framework of the MATLAB \texttt{nnstart} toolbox and particularly useful for noisy data \citep{genccay2001pricing}.} We remark that other training methods may also be used such as Lebenberg-Marquardt \citep{levenberg1944method} and nature-inspired heuristic algorithms such as particle swarm \citep{eberhart1995new}, differential evolution \citep{storn1997differential} or leapfrogging \citep{rhinehart2012leapfrogging}. However, access to gradient information (through the use of a differentiable cost-function) heavily favors the use of gradient-based optimizers. The process of backpropagation, where the errors in the output of the network are propagated back through the network to determine optimal search directions in parameter space is well suited for our particular architecture which utilizes solely one hidden layer. However, a detailed hyperparamater search is beyond the scope of this study since we would like to focus on the performance assessment of the proposed architectures using default options available in \texttt{nnstart} (e.g., mean-squared error loss function). 

For the purpose of tracking the training of the ANN and evaluating its performance, we utilize three different subsets of our overall data with 70\% being utilized for training, and 15\% being utilized for validation and testing each. The training data set is used to adjust the weights of the ANN. The validation data set is used to minimize overfitting, i.e., it is used to verify whether an increase in accuracy over the training data set also yields an increased accuracy over a data set that is not revealed to the network a-priori. The validation data set is used for testing the final solution of the neural network. We note that the input and output features are all scaled between their minimum and maximum values and the scaling parameters are ported for deployment. Within the POD context, the trained ANN can be used as a `blackbox' which takes the present state of the latent space dynamics as the input along with a dimensionless parameter and the present time and uses the memory of its training to determine the time evolution. \textcolor{mycolor}{Although not shown here for brevity, we note that the regression analysis shows a perfect correlation in the time history of the temporal modes and develops a system which could predict their evolution accurately.}  



\subsection{ANN architecture for model order reduction}

In the present study, we devise a framework for developing an ANN in order to create a very fast (equation-free or non-intrusive) integrator/solver for certain types of PDE systems. This idea can be considered as a competitor for POD-GP, with the advantage this is an equation-free and purely data-driven with associated benefits in computational efficiency and model-form uncertainty reduction.

\begin{figure}[!ht]
\centering
\mbox{
\subfigure[Sequential Net]{\includegraphics[height=0.45\textwidth]{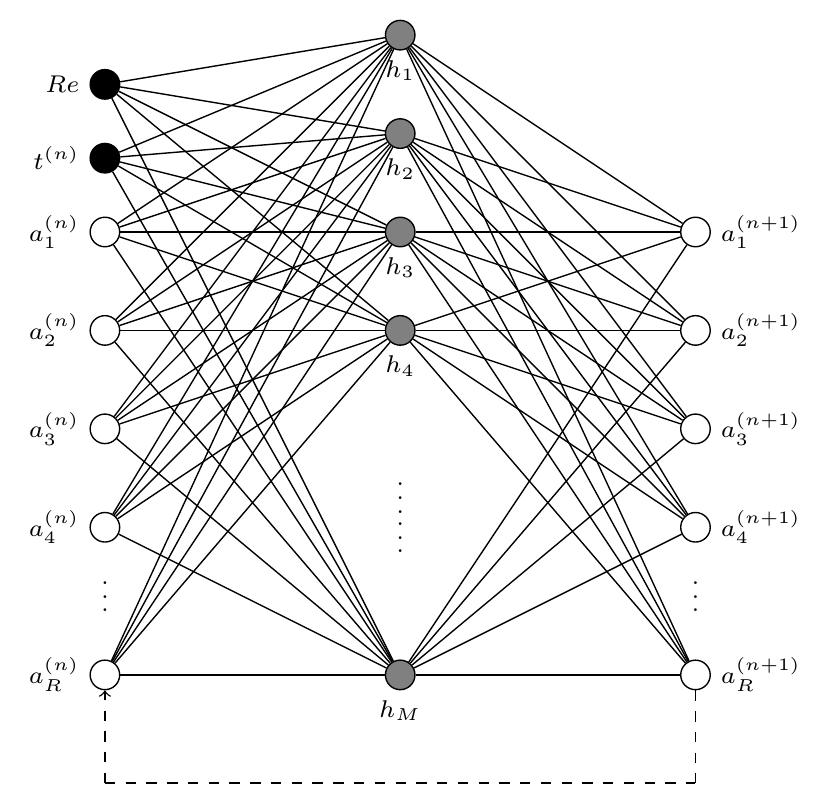}}
\subfigure[Residual Net]{\includegraphics[height=0.45\textwidth]{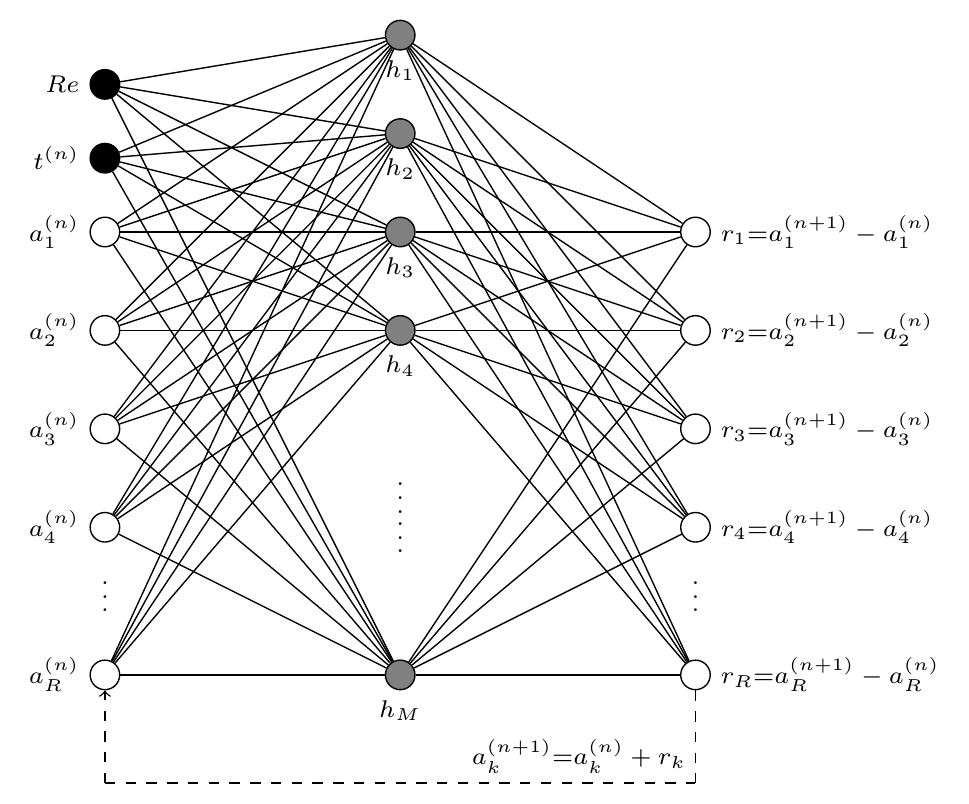}}
}
\caption{The non-intrusive model order reduction enablers for transient flows with the sequential network $\mathbb{M}_1$ (left) and residual network $\mathbb{M}_2$ (right) ANN architectures. We illustrate our architectures for one physical control parameter $Re$ and use only one hidden layer equipped with $M$ neurons and employing $R$ most energetic POD modes.}
\label{fig:burger4}
\end{figure}

Figure~\ref{fig:burger4} illustrates the ANN architectures utilized in this study. 
In these architectures, several inputs are provided, out of which the majority are basis coefficients at the current timestep in the POD-ROM evolution and the remaining two are Reynolds number and time. The training data is also arranged in the same manner. Our outputs are given by just the coefficient values required for a field reconstruction in the next step. The figure referred to above shows the number of inputs (modal POD coefficients, Reynolds number and time) and the number of outputs (modal POD coefficients at the next timestep). The middle layer $l$ contains the neurons with the tan-sigmoid activation function and these are varied in our investigation. Therefore our first mapping approach, sequential network (SN) architecture, equipped by $R=6$ modes can be given by 
\begin{align}
	\mathbb{M}_1 : \{Re,t^{(n)}, a_1^{(n)}, a_2^{(n)}, \hdots, a_{6}^{(n)}\} \in \mathbb{R}^8 \rightarrow \{a_1^{(n+1)}, a_2^{(n+1)}, \hdots, a_{6}^{(n+1)}\} \in \mathbb{R}^6.
\end{align}
\textcolor{mycolor}{We denote this model as POD-ANN-SN. As discussed by \cite{chen2018neural}, machine learning models such as residual networks \citep{he2016deep} define a discrete sequence of finite transformations to a hidden state in the following form \citep{haber2017stable,lu2017beyond}:
\begin{align} \label{eq:ru}
a_{k}^{(n+1)} = a_{k}^{(n)} + r_{k}, \quad \mbox{for} \quad k=1,2, ..., R,
\end{align}
where $r_{k}$ refers to the residual that we can approximate using an ANN. This can be seen as an Euler update in a discrete sense as follows (i.e., defining the residual as $r_{k}=\Delta t f_{k}$)  
\begin{align}
a_{k}^{(n+1)} = a_{k}^{(n)} + \Delta t f_{k},
\end{align}
where $\Delta t = t^{(n+1)}-t^{(n)}$, and $f_{k}$ is the slope
\begin{align}
f_{k} = \frac{a_{k}^{(n+1)} - a_{k}^{(n)}}{\Delta t}. 
\end{align}
To investigate the performance of such a residual network, a similar network shown also in Figure~\ref{fig:burger4} is also designed for $R=6$ largest POD modes, which can be defined as 
\begin{align}
	\mathbb{M}_2 : \{Re,t^{(n)}, a_1^{(n)}, a_2^{(n)}, \hdots, a_{6}^{(n)}\} \in \mathbb{R}^8 \rightarrow \{r_1, r_2, \hdots, r_{6}\} \in \mathbb{R}^6.
\end{align}
and the POD coefficients are then obtained by Eq.~(\ref{eq:ru}) before the consequent step. This model is denoted as POD-ANN-RN in our study. Details related to data acquistion for our network training are provided in Section 7.} We highlight that the main objective of the present work is to test and evaluate the characteristics of the proposed POD-ANN approaches and compare them with the true solution obtained by a projection to the full order model (FOM). Our assessments also includes intrusive results generated by the standard POD-GP framework.

%
%
%

\section{Numerical methods}
\label{sec:num}
In this section, we provide a brief description of the numerical methods employed in this study. First of all, we implement compact difference schemes to approximate the differential operators (both linear and nonlinear terms) in the POD-GP. Compact difference schemes have shown their ability to reach the objectives of high-accuracy and low computational cost for many problems in fluid dynamics \citep{lele1992compact,wang2002analysis}. 

In a compact difference scheme, the first order derivatives can be computed accordingly \citep{lele1992compact}
\begin{equation}\label{eq:cd1}
    \alpha f'_{i-1} + f'_{i} +  \alpha f'_{i+1} = a\frac{f_{i+1}-f_{i-1}}{2\Delta x} + b\frac{f_{i+2}-f_{i-2}}{4\Delta x} ,
\end{equation}
which gives rise to an $\alpha$-family of tridiagonal schemes with $a=\frac{2}{3}(\alpha+2)$, and $b=\frac{1}{3}(4\alpha-1)$. The subscript $i$ represents the spatial grid index in the considered direction, and $\Delta x$ is the uniform grid spacing in that direction. Here, $\alpha=0$ leads to the explicit non-compact fourth-order scheme for first derivative computations. A classical compact fourth-order scheme, which is also known as Pad\'{e} scheme, is obtained by setting $\alpha=1/4$, resulting in $b=0$ with a compact dependence. In computing the linear diffusion term, the second derivative compact scheme is given by
\begin{equation}\label{eq:cd2}
    \alpha f''_{i-1} + f''_{i} +  \alpha f''_{i+1} = a\frac{f_{i+1}-2f_{i}+f_{i-1}}{\Delta x^2} + b\frac{f_{i+2}-2f_{i}+f_{i-2}}{4\Delta x^2} ,
\end{equation}
where $a=\frac{4}{3}(1-\alpha)$, and $b=\frac{1}{3}(10\alpha-1)$. For $\alpha=1/10$ the classical fourth-order Pad\'{e} scheme is obtained. The high-order one-sided derivative formulas are used for the Dirichlet boundary conditions to complete a tridiagonal system of equations for both the first and second order derivative formulas \citep{carpenter1993stable}.


As a time integrator, we use an optimal third-order accurate total variation diminishing Runge-Kutta (TVDRK3) scheme that is given as \citep{gottlieb1998total}
\begin{eqnarray}
u^{(1)}_{i} &=& u^{(n)}_{i} + \Delta t \pounds(u^{(n)}_{i}), \nonumber \\
u^{(2)}_{i} &=& \frac{3}{4} u^{(n)}_{i} + \frac{1}{4} u^{(1)}_{i} + \frac{1}{4}\Delta t \pounds (u^{(1)}_{i}), \nonumber \\
u^{(n+1)}_{i} &=& \frac{1}{3} u^{(n)}_{i} + \frac{2}{3} u^{(2)}_{i} + \frac{2}{3}\Delta t \pounds (u^{(2)}_{i}),
\label{eq:TVDRK}
\end{eqnarray}
where $\pounds (u)$ is the discrete operator of spatial derivatives including linear and nonlinear terms. To perform the inner products defined by Eq.~(\ref{POD_3}), we compute the integral of any arbitrary function $g(x)$ over the domain $\Omega$ by using the Simpson's 1/3 rule \citep{hoffman2001numerical}
\begin{equation}\label{eq:ni1}
    \int_{\Omega} g(x)dx = \frac{\Delta x}{3} \sum_{i=1}^{N_x/2-1} \big( g_{2i} +4g_{2i+1} +g_{2i+2}   \big),
\end{equation}
where $N_x$ is the total number of grid point (e.g., an even number in order to be consistent with the Simpson's integration rule).

\section{Results}

In the following we provide modal coefficient evolution data for our test case and make assessments of the relative performance of the POD-ANN and POD-GP approaches. Although measuring the state evolution error is not equivalent to measuring the error in any standard function space norm or quantity of interest, we found that it simplifies our demonstration and presents an efficient way in our assessments between the proposed approaches and standard POD-GP model. In addition, we also provide details related to the calculation of POD bases and their coefficients for the use of data-driven learning and a-posteriori deployment.


\textcolor{mycolor}{POD bases for the viscous Burgers equation problem are generated by accumulation of snapshot data from the exact solution given by Eq.~(\ref{eq:exact}). As described in Section 3, snapshot sampling is done at different Reynolds numbers between 100 and 1000 at intervals of 100 each (i.e., $Re=100$, $Re=200$, $...$, $Re=1000$). Each choice of a Reynolds number is sampled for 101 snapshots in time between $t=0$ and $t=1$. This leads to a total of 1010 snapshots to construct a snapshot correlation data matrix $C$ with the size of $1010 \times 1010$. Although we generate such 1010 snapshots from the closed form expression, we remark that these snapshots can be obtained from experiments or full order model numerical simulations in most real-life problems.}  The physics of these multiple snapshots for each control parameter is thus spanned by a single set of optimal bases used for subsequent ANN training or Galerkin projection. The ANN training data has been reconstructed with a timestep of $\Delta t = 0.001$ for the same database with 10 different $Re$ from $t=0$ to $t=1$, which leads to 10,010 samples in our training set.  We note that the timestep for POD-GP and POD-ANN deployment are kept identical in our numerical assessments. We highlight that the training of ANN to bypass the Galerkin procedure is an offline cost which utilizes approximately 45 seconds. 

\begin{figure}[!ht]
\centering
\mbox{
\includegraphics[width=0.9\textwidth]{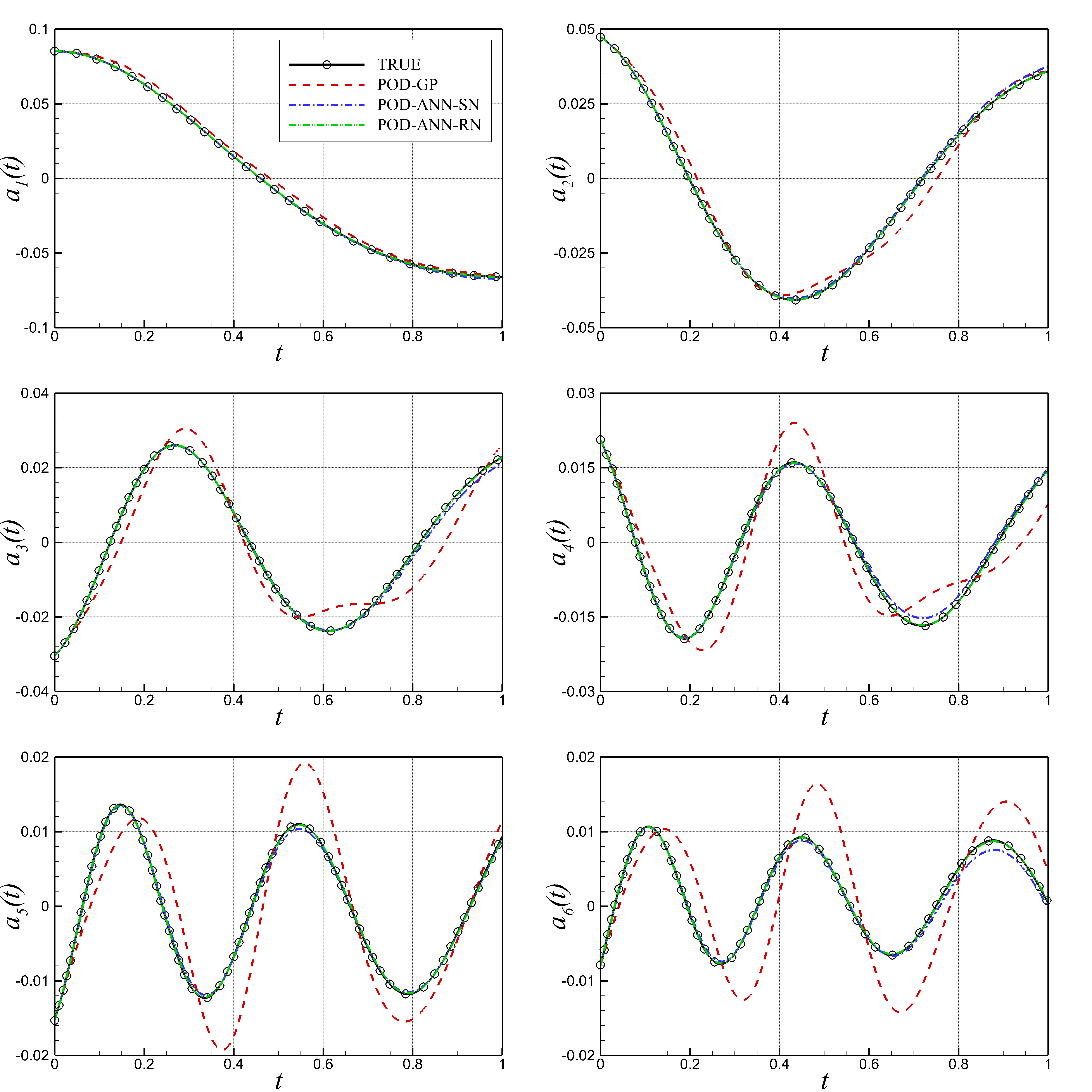}
}
\caption{Numerical assessments of the POD-GP and POD-ANN models with $R=6$ POD modes applied to the Burgers problem for an interpolatory out-of-sample parameter $Re = 750$, where the training set includes snapshots between $Re = 100$ and $Re=1000$. The non-intrusive POD-ANN-SN and POD-ANN-RN results are obtained using $M = 10$ neurons in the hidden layer. Note that the exact solution projected to the reduced order space is shown with the black solid line with circles.}
\label{Fig:burger5}
\end{figure}

\textcolor{mycolor}{We note that the results shown in this subsection are for Reynolds number values which were not within the training dataset range for the ANN. Throughout our study we use a single-layer neural network with $M=10$ neurons and employing $R=6$ POD modes.} Figure~\ref{Fig:burger5} shows a comparison at $Re=750$ on predicting the POD basis coefficients. Our numerical assessment with Reynolds number of 750 represents the interpolative abilities of both POD-GP and POD-ANN approaches. It is clear that the POD-ANN approach preserves state accuracy for a greater duration as compared to the POD-GP approach. At this $Re$, the performance of the both POD-ANN approaches is remarkable. Indeed for the first mode, the POD-ANN approaches preserve the amplitude and phase behavior for the coefficient very well. Coefficients for higher modes tend to show a slight variation in amplitude when compared to DNS data but this variation is quite negligible when compared along the POD-GP approach which shows extensive loss of phase and amplitude at higher modes.

\begin{figure}[!ht]
\centering
\mbox{
\includegraphics[width=0.9\textwidth]{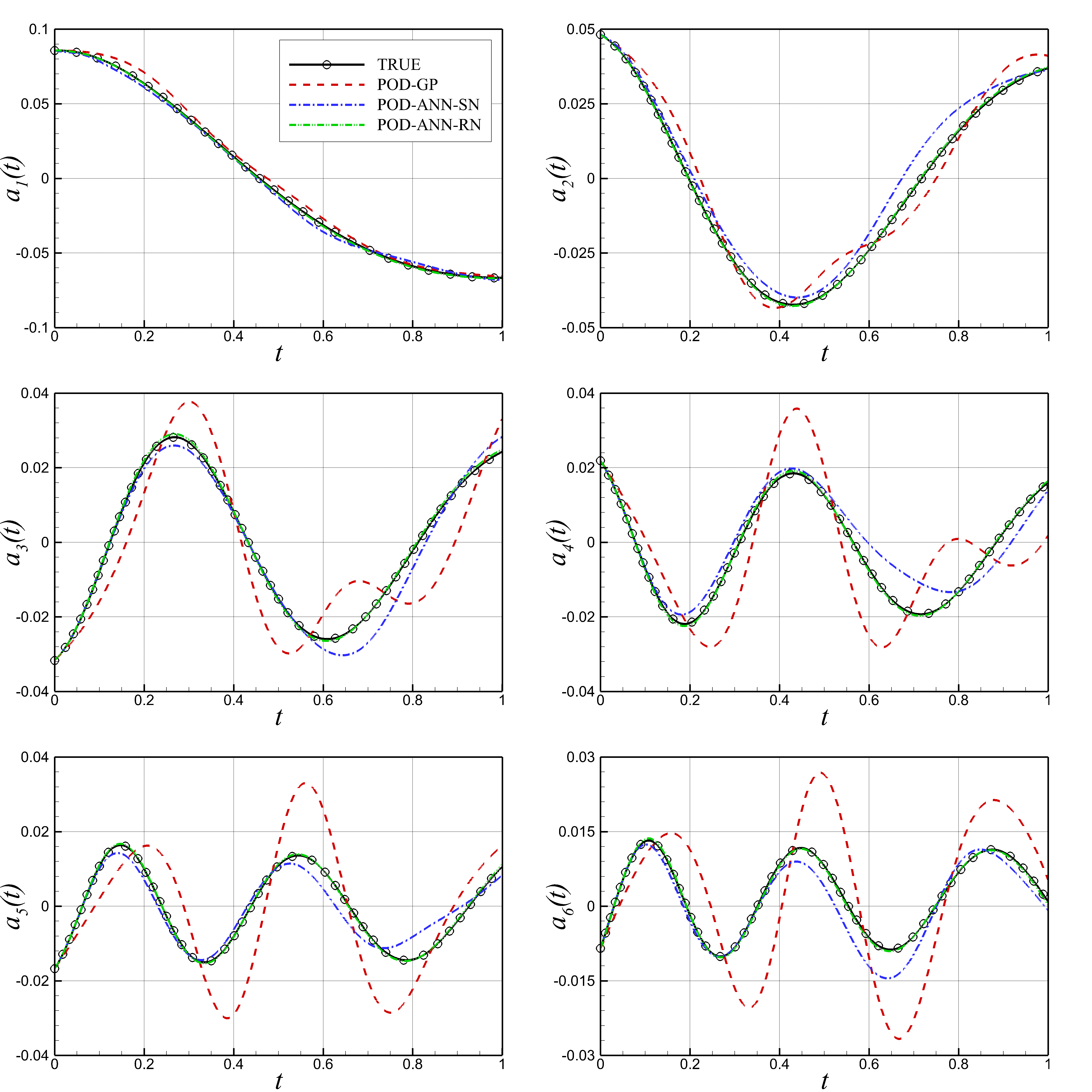}
}
\caption{Numerical assessments of the POD-GP and POD-ANN models with $R=6$ POD modes applied to the Burgers problem at $Re = 1250$, where the training set includes snapshots between $Re = 100$ and $Re=1000$. The non-intrusive POD-ANN-SN and POD-ANN-RN results are obtained using $M = 10$ neurons in the hidden layer. Note that the exact solution projected to the reduced order space is shown with the black solid line with circles.}
\label{Fig:burger6}
\end{figure}

\textcolor{mycolor}{Our next test at $Re=1250$ represents the extrapolative abilities of both POD-GP and POD-ANN models, in which the training data is generated from snapshots between $Re=100$ and $Re=1000$.} Figure~\ref{Fig:burger6} shows a similar study where we have retained only 6 POD modes to predict the flow field at higher $Re$. It can be easily seen that both non-intrusive POD-ANN-SN and POD-ANN-RN approaches yield stable and accurate results. Furthermore, one can immediately notice a significant inaccuracy in the POD-GP approach. \textcolor{mycolor}{Considering only six largest POD modes might not be rich enough to represent the underlying transient behavior.} However, the POD-ANN approaches still perform well in terms of capturing the amplitude of the first six modes even though slight variations in the phase for the highest modes are seen in the $Re=1250$ case as shown in Figure~\ref{Fig:burger6}. \textcolor{mycolor}{Although the POD-GP method shows a considerable deviation from the truth as compared to the POD-ANN approaches, when examining the difference between the POD-ANN-SN and POD-ANN-RN non-intrusive methodologies, we can see that the POD-ANN-RN provides more accurate estimates as compared to the POD-ANN-SN approach which does a decent job at capturing amplitude and phase behavior at this Reynolds number.} 

\begin{figure}[!ht]
\centering
\mbox{
\includegraphics[width=0.9\textwidth]{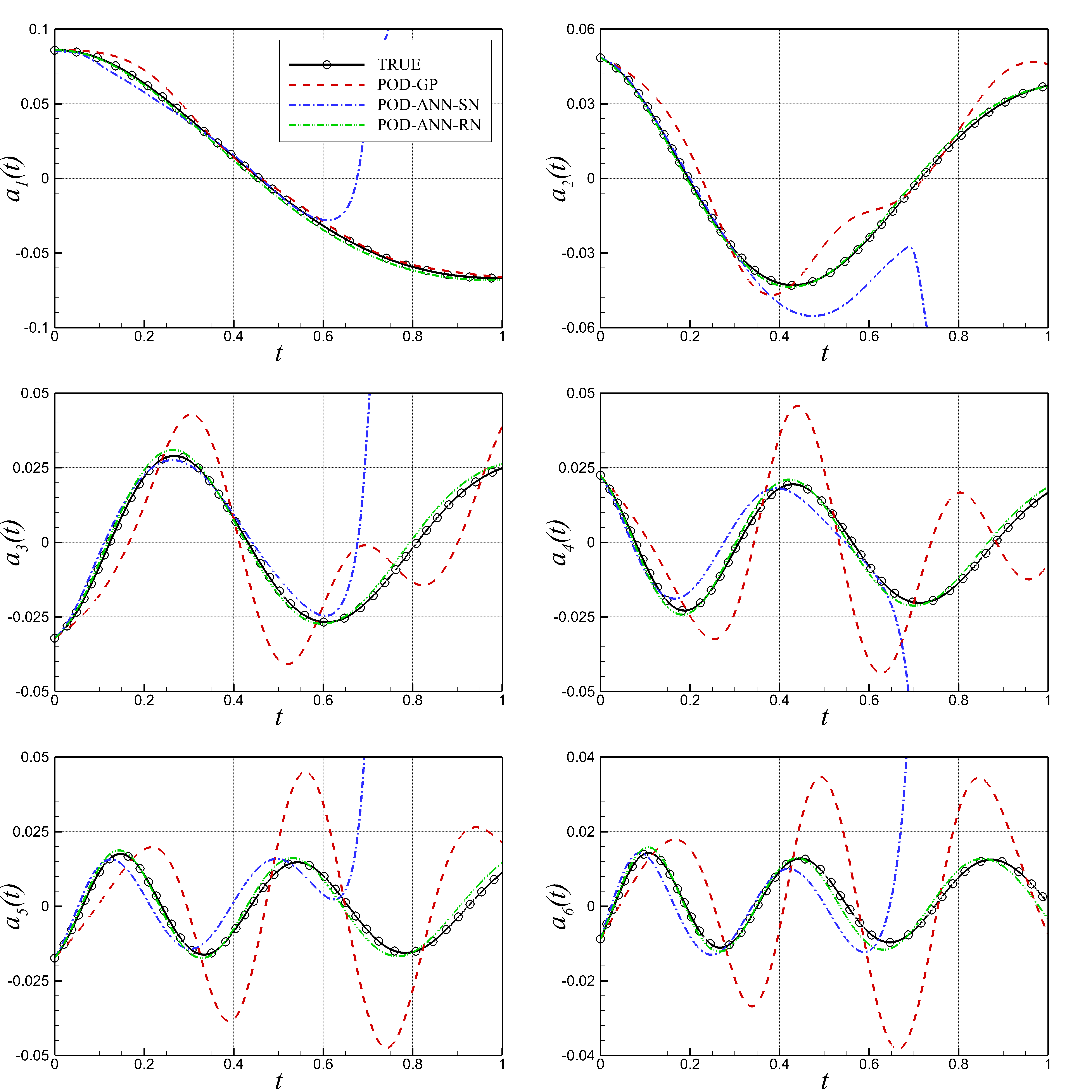}
}
\caption{Numerical assessments of the POD-GP and POD-ANN models with $R=6$ POD modes applied to the Burgers problem at $Re = 2000$, where the training set includes snapshots between $Re = 100$ and $Re=1000$. The non-intrusive POD-ANN-SN and POD-ANN-RN results are obtained using $M = 10$ neurons in the hidden layer. Note that the exact solution projected to the reduced order space is shown with the black solid line with circles.}
\label{Fig:burger7}
\end{figure}

\textcolor{mycolor}{To further investigate the behavior of the proposed ANN architectures at higher Reynolds number, we test the proposed non-intrusive frameworks for Re = 2000. To this end, Figure~\ref{Fig:burger7} presents our numerical assessments at $Re=2000$ that is far beyond the parameter range utilized in the training database. This examination points out that the POD-ANN-RN is more stable than the POD-ANN-SN. This figure also demonstrates significantly more accurate behavior shown by the POD-ANN-RN approach.} In terms of computational costs, the POD-ANN approaches equipped by $M=10$ neurons do not require the use of Runge-Kutta based time stepping and is thus observed to be two orders of magnitude faster than the POD-GP. However, we note that the running time is significantly less than a second for both non-intrusive POD-ANN and intrusive POD-GP approaches with $R=6$ modes. Indeed, we observe that the computational performance of both POD-GP and POD-ANN codes is dominated by data-writing for post-processing.

\textcolor{mycolor}{We stress here that bypassing the computation of the reduced system obtained through Galerkin projection and truncation ensures that our non-intrusive frameworks are \emph{purely} data-driven. The non-intrusive nature of the POD-ANN methodology and the lack of a system of equations that needs to be solved for discretely in time thus leads to an attractive alternative to the POD-GP implementation. Although we illustrate that a robust surrogate model can be obtained by a residual network based ANN architecture without a need to access to the governing equations, we highlight that the intrusive or any other models generated from the first principles allow one to prove important properties like numerical stability, convergence, energy conservation, etc., which are all seemingly lost in blackbox ANN models. However, we refer the readers to \cite{xu2019neural} for an excellent discussion of the key issues in utilizing neural network approaches to inverse problems in differential equations such as universal approximation, regularization, curse of dimensionality and computational efficiency.}

\section{Concluding remarks}

In this study, we have investigated the predictive performance of two different POD-enabled model order reduction approaches for the time evolution of the state. First, we look at the classical Galerkin projection approach where a system of ODEs is developed for the evolution of each state coefficient (corresponding to a scalar field variable). Second, we have developed an equation-free artificial neural network (ANN) framework through data-driven learning which predicts the evolution of the modes without the explicit time stepping approach used in the Galerkin projection. \textcolor{mycolor}{We give a demonstration of the proposed ANN framework for two architectures: (i) a sequential network (SN) where we compute the trajectory of POD coefficients directly, and (ii) a residual network (RN) where we train our model with a discrete residual information, rather than using the sequential state variables.} 
Our numerical assessments are presented for the viscous Burgers equation benchmark problem at various Reynolds number. We notice that the POD-ANN approaches outperforms POD-GP consistently across the different testing environments and proves excellent in both interpolating and extrapolating its learning from obtained data sets for simulations with different control parameters and underlying physics. \textcolor{mycolor}{Furthermore, we demonstrate that POD-ANN-RN yields significantly more stable results especially for the extrapolatory parameter range that are outside of the training range.} We revisit the questions we posed in the introduction to conclude our investigation:
\begin{itemize}
  \item We can conclude with confidence that the POD-ANN approach is superior to the POD-GP approach for the Burgers equation especially for highly truncated systems (i.e., systems represented by a few modes only). 
  \item We can also state that POD-ANN is a viable tool for extrapolation and interpolation beyond the data sets used to train its learning. While it is expected to provide good results within the bounds of the training data set, the Burgers case detailed its capacity to predict the coefficient evolution of different Reynolds numbers in an accurate manner with much greater accuracy than POD-GP.
  \item \textcolor{mycolor}{We find that the residual network formulation is significantly more stable than the sequential network formulation.}
\end{itemize}
Furthermore, we present the following observations which were made during this study:
\begin{itemize}
  \item Once trained, the POD-ANN can be considered to be an `equation-free' ROM architecture.
  \item Although the POD-GP gives us usually better approximations while increasing the retained number modes, the POD-ANN is better suited to more radically truncated systems since an ANN with higher degrees of freedom may show statistical errors on time evolution.
  \item POD-ANN represents a viable framework for modeling highly nonlinear systems when data is available incrementally and constant updates are required. It is also very well suited to situations where model interpretability is not a major concern.
  \item While our training algorithm utilized the standard Bayesian regularization algorithm, it must be noted that there exists many training methodologies which may better preserve the underlying statistical relationships of the inputs and outputs. The effect of training approaches is a future direction of study for this work.
  \item We also recognize that the POD-ANN, despite its obvious benefits, also leads to the drawback of reduced interpretability due to its blackbox nature. In addition, the a-priori requirement of training implies poor generalizability of the POD-ANN framework to completely different physical regimes.
  \item \textcolor{mycolor}{Additionally we note that the use of more advanced regularization techniques in the ANN training (an aspect not considered in this study) would potentially lead to more stable non-intrusive ROM methodologies.} 
  \item Training with noisy data may also be studied for obviating the need for closure modeling in the POD-GP framework.
\end{itemize}

\begin{table}[!h]
\caption{A comparison of intrusive and non-intrusive methodologies describing strengths and weaknesses for each framework.
        \label{tab:2}}
        \smallskip
  \begin{tabular}{| p{0.1in} | p{2.7in} | p{0.1in} | p{2.7in} |}
    \hline
    \multicolumn{2}{|c|}{Intrusive ROMs (POD-GP)} & \multicolumn{2}{c|}{Non-intrusive ROMs (POD-ANN)}\\
    \hline \hline
    + & Solid foundations based on physics and first principles (high interpretability) &     - & Mostly blackbox solvers and packages (low interpretability) \\ \hline
    - & Need access to the governing equations & + & Only need observed/measured data \\ \hline
    + & Generalizes well to new problems with similar physics & - &Poor generalization on unseen problems \\ \hline
	- & Implementations are mostly problem dependent & +  & Extremely convenient for tensor based computing  \\ \hline
	+ & Numerical analysis is easier & -  & Hard to establish rules for its convergence, stability, and energy conservation \\ \hline
    \hline
  \end{tabular}
\end{table}

An overview of our conclusions are provided in Table \ref{tab:2} which contrasts the difference between POD-GP and POD-ANN type methods for ROMs. Although our study utilizes a shallow neural network architecture, we believe that the integration of machine learning tools such as deep learning into the mainstream reduced order modeling ideology would represent a significant increase in its viability for both theoretical and application oriented goals. \textcolor{mycolor}{We would like to conclude that POD-ANN-RN would be a viable ROM procedure for convective flows where the response of complex system dynamics is desired to be represented accurately by a few representative modes.}

\section*{Acknowledgements}
This material is based upon work supported by the U.S. Department of Energy, Office of Science, Office of Advanced Scientific Computing Research under Award Number DE-SC0019290. OS gratefully acknowledges their support. Disclaimer: This report was prepared as an account of work sponsored by an agency of the United States Government. Neither the United States Government nor any agency thereof, nor any of their employees, makes any warranty, express or implied, or assumes any legal liability or responsibility for the accuracy, completeness, or usefulness of any information, apparatus, product, or process disclosed, or represents that its use would not infringe privately owned rights. Reference herein to any specific commercial product, process, or service by trade name, trademark, manufacturer, or otherwise does not necessarily constitute or imply its endorsement, recommendation, or favoring by the United States Government or any agency thereof. The views and opinions of authors expressed herein do not necessarily state or reflect those of the United States Government or any agency thereof.









\bibliographystyle{elsarticle-harv}
\bibliography{references}

\end{document}